\documentclass[a4paper,12pt]{article}
\usepackage{amsmath}
\usepackage{amssymb}
\usepackage{latexsym}
\usepackage{enumerate}
\usepackage{bbm}
\usepackage{amsthm}
\usepackage{tensor}
\usepackage{hyperref}
\newcommand{\be}{\begin{equation}}
\newcommand{\ee}{\end{equation}}

\newtheorem*{theorem}{Theorem}
\newtheorem*{corollary}{Corollary}
\newcommand{\ben}{\begin{equation*}}
\newcommand{\een}{\end{equation*}}
\newcommand{\bean}{\begin{eqnarray*}}
\newcommand{\eean}{\end{eqnarray*}}
\def\bal#1\eal{\begin{align}#1\end{align}}
\newcommand{\ph}{\phantom}
\newcommand{\bsub}{\begin{subequations}}
\newcommand{\esub}{\end{subequations}}

\newcommand{\disfrac}[1][2]{\displaystyle\frac}

\newcommand{\non}{\nonumber}

\newcommand{\pound}{\emph{\textsterling}}
\newcommand{\ima}{\mathbbmtt{i}}
\newcommand{\ts}{\tensor}
\def\bal#1\eal{\begin{align}#1\end{align}}
\newcommand*\xbar[1]{%
  \hbox{%
    \vbox{%
      \hrule height 0.5pt 
      \kern0.5ex
      \hbox{%
        \kern-0.1em
        \ensuremath{#1}%
        \kern-0.1em
      }%
    }%
  }%
}

\begin{document}

\title{\textbf{Lie - point and variational symmetries in minisuperspace Einstein's gravity}}
\vspace{1cm}
\author{\textbf{T. Christodoulakis}\thanks{tchris@phys.uoa.gr}\,, \textbf{N. Dimakis}\thanks{nsdimakis@gmail.com}\,,
\textbf{Petros A. Terzis}\thanks{pterzis@phys.uoa.gr}\\
{\it Nuclear and Particle Physics Section, Physics Department,}\\{\it University of Athens, GR 157--71 Athens}}

\date{}
\maketitle
\begin{center}
\textit{}
\end{center}
\vspace{-1cm}

\abstract{\textit{We consider the application of the theory of symmetries of coupled ordinary differential equations to the case of reparametrisation invariant Lagrangians quadratic in the velocities; such Lagrangians encompass all minisuperspace models. We find that, in order to acquire the maximum number of symmetry generators, one must (a) consider the lapse $N(t)$ among the degrees of freedom and (b) allow the action of the generator on the Lagrangian and/or the equations of motion to produce a multiple of the constraint, rather than strictly zero. The result of this necessary modification of the standard theory (concerning regular systems) is that the Lie - point symmetries of the equations of motion are exactly the variational symmetries (containing the time reparametrisation symmetry) plus the well known scaling symmetry. These variational symmetries are seen to be the simultaneous conformal Killing fields of both the metric and the potential, thus coinciding with the conditional symmetries defined in phase space. In a parametrisation of the lapse for which the potential becomes constant, the generators of the aforementioned symmetries become the Killing fields of the scaled supermetric and the homothetic field respectively.}}

\numberwithin{equation}{section}
\newpage

\section{Introduction}
The theory of variational and/or Lie - point symmetries was initiated by Sophus Lie himself \cite{SLie} and is exploited for quite some time (see textbooks e. g. \cite{Ibra}, \cite{Stephani}, \cite{Olver1}, \cite{Olver2}). The standard theory is concerned with regular systems. In the late nineties and early 2000 the theory was first applied to cosmological minisuperspace models \cite{Marmo}, \cite{Cotsakis1}, \cite{Lamb1}, \cite{Lamb2} and \cite{Cotsakis2}. Since then further applications appeared, for example the recent works: \cite{Vakili}, \cite{Falciano}, \cite{Capoz}, \cite{Sarkar}, \cite{Tsamp1}, \cite{Tsamp2} and \cite{Tsamp3}. In all these investigations the need to reach a regular system was met by gauge fixing the lapse to a specific function (usually $N=1$).

Another way to use the gauge invariance was introduced in the case of Bianchi types \cite{tchris2}, \cite{tchris3}, \cite{tchris4}, \cite{tchris5}. In those cases the lapse was allowed to be defined by the quadratic equation and the automorphisms of the Lie algebra of the specific Bianchi type were identified as Lie - point symmetries of the dynamical equations.

In this work, we consider the lapse as an independent degree of freedom, thereby obtaining also the constraint equation. The influence of such a point of view on the search for symmetry generators is twofold: on the one hand broadens the space of dependent variables, thus increasing the possibility of finding a symmetry; on the other hand introduces more restrictions on the unknown components of the symmetry vector, since their derivatives must satisfy the extra conditions emerging from terms containing $\dot{N}$.

In what follows we consider an action principle,
\be \label{act}
S=\int\!\!L dt,
\ee
corresponding to a singular Lagrangian of the form:
\be \label{Lag}
L=\frac{1}{2N} G_{\kappa\lambda}(q)\dot{q}^\kappa\dot{q}^\lambda-NV(q), \quad\quad \kappa, \lambda =1,\ldots, n.
\ee
with no explicit time dependence. Such Lagrangians are encountered in various cosmological models, where the $q^\kappa$'s represent the scale factor components and/or possible matter fields.

In section 2, we employ the standard theory of variational symmetries, allowing the expansion of the generating symmetry in the $N$ variable. The induced $\omega \frac{\partial L}{\partial N}$ term reproduces the constraint equation, implying that the action of the transformation in $q^\kappa$'s and $t$ keeps invariant \eqref{act} modulo the constraint. The latter differentiates this method from its previous applications in which the action of the symmetry generator is assumed to be exactly zero. Of course, on the solution space, the two requirements can become equivalent: one must simply, even though he has gauge fixed the lapse, allow oneself to interpret the zero as the constraint inserting it by hand.

In the next section, we apply the same way of thinking to the Euler - Lagrange equations ensuing from \eqref{Lag}. In this case the fact that the action of the symmetry generator must be allowed to be a multiple of the Euler - Lagrange equation with respect to $N$ (constraint) is essential. The resulting symmetries are the variational symmetries found in the previous section plus the well known scaling symmetry. We also examine the fate of these symmetries in the particular parametrisation of the lapse, for which the potential is constant, an idea of which the seed is first encountered in \cite{Dewitt}, \cite{Misner}.

In section 4, we show that the found symmetries coincide with the conditional symmetries first defined in \cite{Kuchar} and revisited in \cite{tchris1}.

In section 5 we present a pedagogical application of our results to the Kantowski - Sachs spacetime. Finally, some concluding remarks are included in the discussion.

\section{Symmetries of the action}

A symmetry of the action (or variational symmetry) concerning a regular Lagrangian \cite{Stephani}, \cite{Olver1}, \cite{Olver2} is defined as a transformation $(x,q)\longmapsto(x',q')$ that leaves the action invariant, $\delta S=0$, under the condition that the $q'$'s must remain functions of the $x'$'s. This leads to the well known infinitesimal criterion of invariance
\be \label{cr1}
pr^{(1)} X (L)+ L \frac{d\chi}{dt} = \frac{d f}{dt}
\ee
where $X=\chi(t,q,N) \frac{\partial}{\partial t} + \xi^\alpha(t,q,N) \frac{\partial}{\partial q^\alpha}$ is the generator of the transformation in the space of dependent and independent variables, $pr^{(1)} X$ its first prolongation and $f(t,q,N)$  the so called ``gauge" function. Symmetries found by \eqref{cr1} correspond to existing integrals of motion for the system under consideration.

The usual procedure followed in the literature, due to the fact that \eqref{Lag} is singular, is to gauge fix the lapse function $N$ (usually $N=1$) and then use \eqref{cr1} assuming a notion of pseudo-regularity.  In \cite{tchris1}, we proved how this can lead to the loss of conditional symmetries, introduced in \cite{Kuchar}. These are integrals of motion modulo the constraint, which in the case of \eqref{Lag} is the Hamiltonian itself, $\{Q,\mathcal{H}\}=w(q)\mathcal{H}\approx 0$. Moreover, we exhibited that this problem can be bypassed if one adds in the right hand side of \eqref{cr1}, the constraint equation times $w(q)$.

In what follows, we expand the form of the transformation generated by $X$, by considering $N$ in the same context as the $q^{\kappa}$'s. With the help of \eqref{cr1} we will be led to the conditions that the variational symmetries must satisfy, and acquire their general form for Lagrangians of type $\eqref{Lag}$. We consider the generator of a transformation in the space spanned by $(t,q,N)$ as
\be \label{generator}
X=\chi(t,q,N)\frac{\partial}{\partial t}+ \xi^\alpha(t,q,N)\frac{\partial}{\partial q^\alpha} + \omega(t,q,N)\frac{\partial}{\partial N}
\ee
and its first prolongation
\be
pr^{(1)}X=X+\phi^\alpha\frac{\partial}{\partial \dot{q}^\alpha},
\ee
where there is no $\frac{\partial}{\partial\dot{N}}$ term since the Lagrangian is free of $\dot{N}$. The components $\phi^\alpha$ are given by
\be \label{fa}
\phi^\alpha=\frac{d\xi^\alpha}{dt}-\dot{q}^\alpha \frac{d\chi}{dt} = \xi^\alpha_{,t}-\chi_{,t}\dot{q}^\alpha+\xi^\alpha_{,\beta}\dot{q}^\beta-\chi_{,\beta}\dot{q}^\alpha\dot{q}^\beta +\dot{N}\xi^\alpha_{,0}-\chi_{,0}\dot{N}\dot{q}^\alpha.
\ee

In the appendix \ref{app1}, we present a complete calculation of each term entering \eqref{cr1}. In what follows, for the sake of brevity, we adopt the conventions ``$_{,0}=\frac{\partial}{\partial N}$", ``$_{,\alpha}=\frac{\partial}{\partial q^\alpha}$" and ``$_{,t}=\frac{\partial}{\partial t}$". The next step is to gather the coefficients of the various velocity terms. These must all be identically set to zero, since none of the components of $X$ depends on the $\dot{q}^\alpha$'s and $\dot{N}$.

First we start off from the coefficients of cubic terms $\dot{N}\dot{q}^\kappa\dot{q}^\lambda$ and $\dot{q}^\kappa\dot{q}^\alpha\dot{q}^\beta$:
\bsub
\be
-\frac{\chi_{,0}}{N}G_{\kappa\lambda}+\frac{\chi_{,0}}{2N}G_{\kappa\lambda}=0 \Rightarrow -\frac{\chi_{,0}}{2N}G_{\kappa\lambda} =0 \Rightarrow \chi_{,0}=0 \Rightarrow \chi=\chi(t,q)
\ee
\be \label{xit}
-\frac{\chi_{,\beta}}{N}G_{\kappa\alpha}+\frac{\chi_{,\beta}}{2N}G_{\kappa\alpha}=0 \Rightarrow -\frac{\chi_{,\beta}}{2N}G_{\kappa\alpha} =0 \Rightarrow \chi_{,\beta}=0\Rightarrow \chi=\chi(t).
\ee
\esub
We continue with the coefficients of quadratic terms $\dot{N}\dot{q}^\kappa$ and $\dot{q}^\kappa \dot{q}^\lambda$:
\bsub
\bal \label{etatq}
\frac{\xi^\alpha_{,0}}{N} G_{\kappa\alpha} \Rightarrow \xi^\alpha_{,0}=0\Rightarrow \xi^\alpha=\xi^\alpha(t,q)\\
\frac{1}{2 N} \left(\xi^\alpha G_{\kappa\lambda,\alpha}+\xi^\alpha_{,\kappa}G_{\alpha\lambda}+ \xi^\alpha_{,\lambda}G_{\alpha\kappa}\right) -\frac{\chi_{,t}}{N}G_{\kappa\lambda}+\frac{\chi_{,t}}{2N}G_{\kappa\lambda}-\frac{\omega}{2N^2}G_{\kappa\lambda}&=0 \Rightarrow \non\\ \label{final}
\frac{1}{2 N} \left(\xi^\alpha G_{\kappa\lambda,\alpha}+\xi^\alpha_{,\kappa}G_{\alpha\lambda}+ \xi^\alpha_{,\lambda}G_{\alpha\kappa}\right) -\frac{\chi_{,t}}{2N}G_{\kappa\lambda}-\frac{\omega}{2N^2}G_{\kappa\lambda}&=0.
\eal
\esub
From the coefficients of the linear in $\dot{N}$ and $\dot{q}^\kappa$ terms, we get:
\bsub
\be \label{ftq}
\chi_{,0} N V-f_{,0}=0\Rightarrow f_{,0} =0 \Rightarrow f=f(t,q)
\ee
\be
\frac{\xi^\alpha_{,t}}{N}G_{\kappa\alpha} - \chi_{,\kappa}NV -f_{,\kappa} =0 \Rightarrow \frac{\xi^\alpha_{,t}}{N}G_{\kappa\alpha} -f_{,\kappa} =0.
\ee
\esub
The latter equation, on account of \eqref{etatq} and \eqref{ftq}, leads to $\xi^\alpha_{,t}=0$ and $f_{,\kappa}=0$, that is $\xi^\alpha=\xi^\alpha(q)$ and $f=f(t)$.

Finally we are left with an equation that is composed of the zero order terms in the velocities
\be
-N\xi^\alpha V_{,\alpha}-\omega V - \chi_{,t} N V - f_{,t} =0
\ee
which can be solved for $\omega$ if $V\neq 0$,
\be
 \omega =- N\xi^\alpha \frac{V_{,\alpha}}{V}- N\chi_{,t}- \frac{f_{,t}}{V}.
\ee
By substitution to \eqref{final}, which is the only remaining unsolved equation,  we get
\be
\pound_\xi G_{\kappa\lambda} = \left(-\xi^\alpha \frac{V_{,\alpha}}{V}-\frac{f_{,t}}{NV}\right)G_{\kappa\lambda}
\ee
and since $\xi^\alpha$ is $t$ and $N$ independent, the only possibility for the gauge function $f$ is to be constant. Thus we are left with the following conditions for the vector $\xi:=\xi^\alpha\frac{\partial}{\partial q^\alpha}$ defined on the part of the configuration space spanned by $q^\alpha$'s:
\be \label{varcondition}
\pound_\xi G_{\kappa\lambda}=\tau(q) G_{\kappa\lambda}
\ee
where
\be \label{liev}
\tau(q)=-\xi^\alpha \frac{V_{,\alpha}}{V}\Rightarrow \pound_\xi V = -\tau(q) V.
\ee
Therefore the most general generator for the variational symmetries of \eqref{Lag} is
\bal \label{vargen}
X&=X_1+X_2\non\\
X_1&=  \xi^\alpha(q)\frac{\partial}{\partial q^\alpha} - \tau(q)\,N\frac{\partial}{\partial N},\, X_2=\chi(t)\frac{\partial}{\partial t} +\chi(t)_{,t}N\frac{\partial}{\partial N}
\eal

\begin{theorem}
The variational symmetries for the action \eqref{act} have an infitestimal generator of the form \eqref{vargen} provided that
\be
\pound_\xi G_{\mu\nu}= \tau(q) G_{\mu\nu} \quad\quad and \quad\quad \pound_\xi V= - \tau(q) V.
\ee
\end{theorem}

It is noteworthy that $\chi(t)$ remains an unrestricted function of time, a fact that reflects the time reparametrisation invariance of the theory stemming out of the singular Lagrangian \eqref{Lag}. Equations \eqref{varcondition} and \eqref{liev} signify that, in order to have a variational symmetry for Lagrangians of this form, the vector $\xi$ must be a simultaneous conformal Killing vector of both the potential and the super-metric with conformal factors of opposite signs, in complete accordance with the results exhibited in \cite{tchris1} in the context of conditional symmetries. Note that this final form of the above generator \eqref{vargen} is not affected even if we assume a zero potential $V=0$.

The integral of motion that corresponds to the invariance transformation generated by \eqref{vargen} is
\begin{align} \nonumber
Q&= \xi^\alpha \frac{\partial L}{\partial \dot{q}^\alpha}+ \chi \, L- \chi \dot{q}^\alpha \frac{\partial L}{\partial \dot{q}^\alpha} \\ \nonumber
&= \xi^\alpha \frac{1}{N}G_{\kappa\alpha}\dot{q}^\kappa - \chi(t) \left(\frac{1}{2N}G_{\kappa\lambda}\dot{q}^\kappa\dot{q}^\lambda+NV\right) \\
&= \xi^\alpha \frac{1}{N}G_{\kappa\alpha}\dot{q}^\kappa - \chi(t) E^0, \label{intq}
\end{align}
where $E^0$ the Euler - Lagrange equation which corresponds to $N$. If we express the integral \eqref{intq} in the phase space ($\dot{q}^\alpha \mapsto p_\alpha =\frac{\partial L}{\partial \dot{q}^\alpha}=\frac{1}{N}G_{\mu\alpha}\dot{q}^\mu$), with a Hamiltonian function
\be
H =N\mathcal{H}= N\left(\frac{1}{2}G^{\kappa\lambda} \pi_\kappa \pi_\lambda + V\right)
\ee
and first class constraints
\be
\pi_N \approx 0 \quad, \quad \mathcal{H}\approx 0,
\ee
we get
\be \label{intofmo}
Q=\xi^\alpha p_\alpha -\chi(t)N \mathcal{H} \approx \xi^\alpha \pi_\alpha.
\ee
As one can easily check $Q$ is a conditional symmetry, i.e. it is an integral of motion due to the constraint $\mathcal{H}\approx 0$
\begin{align} \nonumber
\frac{d Q}{dt}& =\{\xi^\alpha \pi_\alpha, H+u^N \pi_N\}- \frac{\partial}{\partial t}(\chi(t)N \mathcal{H})-\chi(t)\{N\mathcal{H},H+u^N\,\pi_N\}  \\
&=- \tau(q) N \mathcal{H} - \chi_{,t} N \mathcal{H} - \chi\, u^N \mathcal{H} \approx 0.
\end{align}
The $\chi(t)N \mathcal{H}$ term in \eqref{intofmo} is more or less trivial in the sense that the Hamiltonian is weakly zero, but it is very interesting that the generalization of the generator $X$, by admitting transformations in the variable $N$, led to the freedom of time reparametrisation through $\chi(t)$ and the scaling of the lapse function in the component of $\partial_N$.

\section{Lie point symmetries of the Euler - Lagrange Equations}
The Euler - Lagrange equations for Lagrangian \eqref{Lag} are
\begin{align} \label{ELN}
\frac{\partial L}{\partial N} &= 0 \\ \label{ELq}
\frac{\partial L}{\partial q^\kappa}-\frac{d}{dt}\left(\frac{\partial L}{\partial \dot{q}^\kappa}\right) &= 0 .
\end{align}
For a valid Lagrangian, the $n$ equations \eqref{ELq} correspond to the spatial set of Einstein's equations, while \eqref{ELN} represents the quadratic constraint equation involving only the velocities (i.e. the $G^0_0=T^0_0$ Einstein equation). These two sets lead respectively to
\begin{subequations} \label{eul}
\be \label{con}
E^0 := G_{\mu\nu}\dot{q}^\mu \dot{q}^\nu + 2N^2 V =0
\ee
\be \label{spa}
E^\kappa := \ddot{q}^\kappa + \Gamma^\kappa_{\mu\nu}\dot{q}^\mu \dot{q}^\nu - \frac{\dot{N}}{N} \dot{q}^\kappa + N^2 V^{,\kappa}=0,
\ee
\end{subequations}
where $V^{,\kappa}=G^{\rho\kappa} V_{,\rho}$.

Again, the usual treatment for the search of Lie - point symmetries is to gauge fix the lapse (say, $N=1$) and then use the infitestimal criterion $pr^{(2)}X(E^\kappa)=0$, $\mathrm{mod}\; E^\kappa=0$, with $pr^{(2)}X$ being the second prolongation of a generator whose coefficients depend only on $t$ and $q$. In order to work without gauge fixing and exploit the freedom introduced by the invariance of our theory, we choose to expand the generating transformation as in the previous section, see equation \eqref{generator}. Moreover we make use of the constraint equation, demanding
\begin{subequations} \label{newcrit}
\be\label{newcrit1}
pr^{(1)}X(E^0) = T(t,q,N) E^0
\ee
\be \label{newcrit2}
pr^{(2)}X(E^\kappa) = \left(P_{1\; \alpha}^\kappa(t,q,N) \dot{q}^\alpha+ P_2^\kappa(t,q,N) \dot{N} + P^\kappa(t,q,N) \right) E^0, \;\mathrm{mod} \; E^\kappa =0.
\ee
\end{subequations}
This necessary modification of the infinitesimal criterion can be seen as an interpretation of the zero in terms of the constraint, which indeed vanishes on the solution space. The parenthesis in the right hand side of \eqref{newcrit2} is the only existing possibility, since after the replacement of the accelerations from \eqref{spa}, the left hand side contains terms at most cubic in the velocities.

The second prolongation of $X$ is
\be \label{pro2}
pr^{(2)}X = X + \phi^\alpha \frac{\partial}{\partial\dot{q}^\alpha} + \Omega \frac{\partial}{\partial\dot N} + \Phi^\alpha \frac{\partial}{\partial \ddot{q}^\alpha}
\ee
where $\partial_{\ddot{N}}$ has been omitted on account of $\ddot{N}$ being absent from $E^0$ and $E^\kappa$, $\phi^\alpha$ is the same as before (see equation \eqref{fa}),
\be
\Omega = \frac{d \omega}{d t} - \dot{N} \frac{d \chi}{dt}=\omega_{,t} + \dot{N}(\omega_{,0}-\chi_{,t})+\omega_{,\beta}\dot{q}^\beta- \chi_{,\beta}\dot{N}\dot{q}^\beta- \chi_{,0}\dot{N}^2
\ee
and
\begin{align} \nonumber
\Phi^\alpha &=\frac{d\phi^\alpha}{dt}-\ddot{q}^\alpha \frac{d\chi}{dt}  \\ \nonumber
&= \xi^\alpha_{,tt}+(2\xi^\alpha_{,\beta t}-\chi_{,tt}\delta^\alpha_\beta)\dot{q}^\beta+ 2\dot{N} \xi^\alpha_{,0t}+ \left(\xi^\alpha_{,\beta\gamma}-\chi_{,\beta t}\delta^\alpha_\gamma-\chi_{,\gamma t}\delta^\alpha_\beta\right)\dot{q}^\beta\dot{q}^\gamma \\ \nonumber
& + 2\left(\xi^\alpha_{,0\beta}-\chi_{,0t}\delta^\alpha_\beta\right)\dot{q}^\beta\dot{N}+\dot{N}^2\xi^\alpha_{,00}- \chi_{,\beta\gamma}\dot{q}^\alpha\dot{q}^\beta\dot{q}^\gamma- 2\chi_{,0\beta}\dot{N}\dot{q}^\alpha\dot{q}^\beta- 2\chi_{,00}\dot{N}^2\dot{q}^\alpha \\
&+ \left(\xi^\alpha_{,\beta}-2\chi_{,t}\delta^\alpha_\beta\right)\ddot{q}^\beta- 2\chi_{,\beta}\ddot{q}^\alpha\dot{q}^\beta-\chi_{,\beta}\ddot{q}^\beta\dot{q}^\alpha+\ddot{N}\xi^\alpha_{,0}- \chi_{,0}\dot{N}\ddot{q}^\alpha- \chi_{,0}\ddot{N}\dot{q}^\alpha.
\end{align}
In appendix \ref{app2} we give the results of the action of each term of the prolongation on equations $E^\kappa$, here we gather the terms regarding $\ddot{N}$, $\dot{N}$ and $\dot{q}$'s (after the substitution of $\ddot{q}^\kappa$'s from \eqref{spa}). As previously stated their coefficients must be zero, since none of the entailing unknown functions has a dependence in the velocities. For equation \eqref{newcrit2} we get the following coefficients concerning the acceleration $\ddot{N}$:
\bsub
\be
\chi_{,0}=0 \Rightarrow \chi=\chi(q,t)
\ee
\be
\xi^\kappa_{,0}=0 \Rightarrow \xi^\kappa=\xi^\kappa(q,t).
\ee
\esub
We proceed with the coefficients of the cubic terms $\dot{N}\dot{q}^\mu\dot{q}^\nu$:
\be \label{ena}
P_2^\kappa\, G_{\mu\nu}+ \frac{3}{2N}\left(\chi_{,\mu}\delta^\kappa_\nu+ \chi_{,\nu} \delta^\kappa_\mu\right) =0 \Rightarrow P_2^\kappa =\frac{F^\kappa(t,q)}{N}
\ee
where we have set $F^\kappa(t,q)=-\frac{3}{2\, n}G^{\rho\sigma}\left(\chi_{,\rho}\delta^\kappa_\sigma+ \chi_{,\sigma} \delta^\kappa_\rho\right)$.
At this point it is easier to consider the coefficient of the linear term $\dot{N}$
\be
\frac{\xi^\kappa_{,t}}{N}+ 2N^2\, V\, P_2^\kappa=0 \Rightarrow \frac{\xi^\kappa_{,t}}{N}+ 2 N \, V F^\kappa(t,q) =0,
\ee
which leads to
\be
\xi^\kappa_{,t}=0\Rightarrow \xi^\kappa=\xi^\kappa(q) \quad \text{���} \quad F^\kappa=0\Rightarrow P_2^\kappa =0.
\ee
By virtue of \eqref{ena}, we obtain $\chi_{,(\alpha}\delta^\kappa_{\beta)}=0$ and by contracting $\kappa$, $\beta$ we arrive to
\be
\chi_{,\alpha}=0\Rightarrow \chi=\chi(t).
\ee
Further on, we take up successively the coefficients of the terms $\dot{N}\dot{q}^\kappa$, $\dot{q}^\alpha\dot{q}^\mu\dot{q}^\nu$, $\dot{q}^\kappa$ and $\dot{q}^\mu\dot{q}^\nu$:
\bsub
\bal
\frac{1}{N}\left(\frac{\omega}{N}-\omega_{,0}\right)=0 \Rightarrow \omega= N\tilde{\omega}(t,q)\\
P^\kappa_{1\; (\alpha}G_{\mu\nu)}=0 \Rightarrow P^\kappa_{1\; \alpha}=0\\
\frac{\omega_{,t}}{N}+\chi_{,tt}=0 \Rightarrow \tilde{\omega}_{,t}= - \chi_{,tt}\Rightarrow \tilde{\omega}=-\chi_{,t}+ h(q) \Rightarrow \omega=N(h(q)-\chi_{,t})\\
\label{liegam}
\pound_\xi \Gamma^\kappa_{\mu\nu}- \frac{1}{2}\left(h_{,\mu}\delta^\kappa_\nu+ h_{,\nu} \delta^\kappa_\mu\right)-P^\kappa G_{\mu\nu}=0
\eal
\esub
and finally we are left with the zero order terms in the velocities, which lead to
\begin{align} \nonumber
&-N^2\, V^{,\alpha} \left(\xi^\kappa_{,\alpha}-2\chi_{,t}\delta^\kappa_\alpha\right) + 2\, \omega \, N \, V^{,\kappa} + \xi^\alpha\, N^2 \, (V^{,\kappa})_{,\alpha} - 2 \, N^2\, P^\kappa \, V =0  \\ \label{pkap}
&\Rightarrow P^\kappa = \frac{\pound_\xi V^{,\kappa}}{2V} + \frac{V^{,\kappa}}{V}\, h(q).
\end{align}
Equations \eqref{liegam} and \eqref{pkap} are the final set of equations that have to be satisfied in order for \eqref{newcrit2} to hold. We proceed in the same manner in order to gain further conditions from \eqref{newcrit1}.
Due to the restrictions already imposed on the functions entering the generator $X$, i.e. $\chi=\chi(t)$, $\xi^\alpha=\xi^\alpha(q)$ and $\omega=N(h(q)-\chi_{,t})$, the action of $pr^{(1)}X$ on $E^0$ becomes
\be
pr^{(1)}X(E^0)=\left(\pound_\xi G_{\mu\nu}-2\chi_{,t} G_{\mu\nu}\right)\dot{q}^\mu \dot{q}^\nu + 2N^2 \left(\pound_\xi V+2h(q)V-2\chi_{,t} V\right)
\ee
and it must equal
\be
T(t,q,N)E^0=T(t,q,N)G_{\mu\nu}\dot{q}^\mu\dot{q}^\nu + 2N^2 T(t,q,N) V
\ee
which means that $T$ must necessarily be of the form, $T(t,q,N)=\tau(q)+\tilde{\tau}(t)$. This leads to the subsequent set of equations
\begin{align} \label{confg}
\pound_\xi G_{\mu\nu}=\tau G_{\mu\nu} \\ \label{confv}
\pound_\xi V = (\tau - 2 h) V \\
\tilde{\tau} = -2 \chi_{,t}.
\end{align}

A connection can be established between the functions $h(q)$ and $\tau(q)$, because whenever \eqref{confg} holds, the action of the Lie derivative on the Christoffel symbols is
\begin{align} \nonumber
\pound_\xi \Gamma^\kappa_{\mu\nu} &\equiv \frac{1}{2}G^{\kappa\rho}\Big( (\pound_\xi G_{\rho\mu})_{;\nu}+ (\pound_\xi G_{\rho\nu})_{;\mu}- (\pound_\xi G_{\mu\nu})_{;\rho} \Big) \\
\label{confgamma} &= \frac{1}{2} \left(\tau_{,\mu}\delta^\kappa_\nu+ \tau_{,\nu}\delta^\kappa_\mu - \tau^{,\kappa} G_{\mu\nu}\right).
\end{align}
where \eqref{confg} has been used. By comparing \eqref{confgamma} with \eqref{liegam} and using \eqref{pkap} one is led, for $n>1$, to
\be
h(q)=\tau(q) +c,
\ee
where $c$ is a constant. The complete proof is given in appendix \ref{proof}.

Therefore, the infinitesimal generator of the Lie point symmetries of equations \eqref{eul} is
\be \label{liegen}
X= \chi(t)\frac{\partial}{\partial t} + \xi^\alpha(q)\frac{\partial}{\partial q^\alpha} - N\left(\tau(q)+ c -\chi_{,t}\right)\frac{\partial}{\partial N}
\ee
with $\chi(t)$ remaining an arbitrary function of time and $\xi^\alpha(q)$, $\tau$ being specified by
\begin{subequations} \label{confhom}
\be\label{lieg1}
\pound_\xi G_{\mu\nu}=\tau G_{\mu\nu}
\ee
\be\label{liev1}
\pound_\xi V = -(\tau +2 c) V.
\ee
\end{subequations}
The above equations are identical with \eqref{varcondition}, \eqref{liev} when the constant $c$ is zero.

In order to gain some insight for the presence of $c$, we will consider for a moment the case where equations \eqref{con} and \eqref{spa} represent Einstein's equations in vacuum. In this case, Lagrangian \eqref{Lag} is understood to be identified with
\be
L = \frac{1}{2\, N} G^{ijkl} \dot{\gamma}_{ij}\dot{\gamma}_{ij} -N\, \sqrt{\gamma} \, \tensor[^{\scriptscriptstyle{(d)}}]{R}{},
\ee
where $\gamma_{ij}$ is the scale factor matrix of the homogeneous $d$ dimensional space, $\tensor[^{\scriptscriptstyle{(d)}}]{R}{}$ is the scalar curvature of this space and $G^{ijkl}=\sqrt{\gamma}\left(\gamma^{ik}\gamma^{jl}+ \gamma^{il}\gamma^{jk}- 2\gamma^{ij}\gamma^{kl}\right)$. As it is known $G^{ijkl}$ is a homogeneous function of degree $\frac{n}{2}-2$ in the $\gamma_{ij}$'s. The potential term, $\sqrt{\gamma} \, \tensor[^{\scriptscriptstyle{(d)}}]{R}{}$, is also homogeneous of degree $\frac{n}{2}-1$, where $n=\frac{d\,(d+1)}{2}$ for a full scale factor. This, according to Euler's theorem for homogeneous functions implies that
\bal \nonumber
q^\alpha\, G_{\mu\nu,\alpha}= \left(\frac{n}{2}-2\right)\, G_{\mu\nu}\quad &\text{and} \quad q^\alpha\, V_{,\alpha}= \left(\frac{n}{2}-1\right)\, V \Rightarrow \\ \label{euler}
\pound_y G_{\mu\nu} =\frac{n}{2}\, G_{\mu\nu} \quad &\text{and} \quad \pound_y V= \left(\frac{n}{2}-1\right)\, V
\eal
where $y:=q^\alpha\frac{\partial}{\partial q^\alpha}$ and the correspondence $G^{ijkl}\rightarrow G_{\mu\nu}$, $\gamma_{ij}\rightarrow q^\alpha$, and $\sqrt{\gamma} \, \tensor[^{\scriptscriptstyle{(3)}}]{R}{}\rightarrow V$ has been utilized.
Due to \eqref{euler} we are led to redefine the component $\xi^\alpha$, $\tau$ of the generator as follows:
\bal
\xi^\alpha=\underline{\xi}^\alpha+\frac{2c}{1-n}\,q^\alpha,\quad \tau=\underline{\tau}+\frac{n\, c}{1-n}\, .
\eal
Under this redefinition equations \eqref{confhom}, by virtue of \eqref{euler}, transform to
\bsub \label{xiunder}
\bal
\pound_{\underline{\xi}}G_{\mu\nu}&=\underline{\tau}G_{\mu\nu}\\
\pound_{\underline{\xi}}V&=-\underline{\tau}V \,.
\eal
\esub
These equations suggest that the generator\eqref{liegen} is decomposed into $X=X_1+X_2-\frac{2\,c}{1-n}\,Y$, where
\bsub
\bal \label{gen1}
X_1&=\underline{\xi}^\alpha(q)\frac{\partial}{\partial q^\alpha} - \underline{\tau}(q)\, N\, \frac{\partial}{\partial N}\\ \label{gen2}
X_2&=\chi(t)\frac{\partial}{\partial t} +\chi(t)_{,t}\, N\, \frac{\partial}{\partial N} \\ \label{genY}
Y&=q^\alpha\frac{\partial}{\partial q^\alpha}+\frac{1}{2} N \frac{\partial}{\partial N}.
\eal
\esub
Thus, we conclude that $Y$ is indeed the well known scaling symmetry of vacuum Einstein's equations \cite{Stephani}.

The $X_1$ represents the possible existing conditional symmetries (defined in the phase space) encoded in the simultaneous conformal Killing fields of the potential $V$ and the configuration space metric $G_{\mu\nu}$. The $X_2$ represents the time reparametrisation invariance encoded in the arbitrary function $\chi(t)$.

All the above considerations lead to the following theorem:

\begin{theorem}
The Lie - point symmetries for equations \eqref{eul} in pure minisuperspace gravity are the ones with the infitestimal generator $X_1$ which satisfy
\be
\pound_\xi G_{\mu\nu}= \underline{\tau}(q) G_{\mu\nu} \quad\quad and \quad\quad \pound_\xi V= - \underline{\tau}(q) V
\ee
plus the scaling symmetry generator $Y$ and the time reparametrisation generator $X_2$.
\end{theorem}

In \cite{tchris1} the idea of choosing a lapse parametrization $\xbar{N}= N\, V$ so that the potential becomes $q^\alpha$ independent was first put in practical use: as it is known, the theory is insensitive to a scaling of the lapse function $N$ due to the reparametrisation transformations of the independent and dependent variables
\begin{equation*}
t=f(\tilde{t}),\quad N(t) \rightarrow \tilde{N}(\tilde{t}) := N(f(\tilde{t}))\,f'(\tilde{t}), \quad q^\alpha(t) \rightarrow \tilde{q}^\alpha(\tilde{t}) := q^\alpha(f(\tilde{t})).
\end{equation*}
which can easily be seen to leave the action form invariant. Under this change of $N$ the scaled Lagrangian becomes
\be \label{phlag}
\xbar{L} = \frac{1}{2 \xbar{N}}\, \xbar{G}_{\kappa\lambda}\,\dot{q}^\kappa \dot{q}^\lambda - \xbar{N}
\ee
with $\xbar{G}_{\kappa\lambda}= V G_{\kappa\lambda}$ and trivially $\xbar{V}=1$. For a vector $\underline{\xi} = \underline{\xi}^\alpha \frac{\partial}{\partial q^\alpha}$ associated to the generator $X_1$ and thus satisfying \eqref{xiunder}, we have
\begin{align} \nonumber
\pound_{\underline{\xi}} \xbar{G}_{\mu\nu} &= \pound_{\underline{\xi}} (V\, G_{\mu\nu}) \non \\
& = G_{\mu\nu}\pound_{\underline{\xi}} V + V\pound_{\underline{\xi}} G_{\mu\nu} \non \\ \non
& = -\tau\, V\, G_{\mu\nu}+\tau\, V\, G_{\mu\nu} \\
&=0
\end{align}
and of course trivially $\pound_{\underline{\xi}} \xbar{V} = 0$. This means that the simultaneous conformal Killing fields $\underline{\xi}$ are becoming Killing fields of both $\xbar{G}_{\kappa\lambda}$ and $\xbar{V}$. It is very interesting that the well known scaling generator $Y$ becomes just the homothetic Killing field of this metric
\bal
\pound_Y\xbar{G}_{\mu\nu}&= \pound_Y (V\, G_{\mu\nu}) \non \\
& = G_{\mu\nu}\pound_Y V + V\pound_Y G_{\mu\nu} \non \\ \non
& = V\, \frac{n}{2}\,G_{\mu\nu} \\
&=\frac{n}{2}\,\xbar{G}_{\mu\nu}
\eal
leading to a sort of integral of motion as explained in the next section.

Due to the previous considerations we can state the following
\begin{theorem}
The Lie - point symmetries of equations \eqref{eul} are either Killing fields or a homothecy of the scaled supermetric $\xbar{G}_{\mu\nu}=V  G_{\mu\nu}$ plus the time reparametrisation generated by $X_2$.
\end{theorem}
with the further implication
\begin{corollary}
The maximum number of Lie - point symmetries of equations \eqref{eul} is
\begin{equation*}
\frac{n(n+1)}{2}+2,
\end{equation*}
\end{corollary}
i. e. the maximum possible number of Killing fields plus the homothetic field, plus the reparametrisation generator $X_2$. 

\section{Conditional Symmetries and phase space description}

The transition to the Hamiltonian description, for the scaled Lagrangian \eqref{phlag}, is achieved with the help of the momenta
\bal\label{pa}
\pi_\alpha=\frac{\partial \xbar{L}}{\partial \dot{q}^\alpha}\Rightarrow \pi_\alpha=\frac{1}{\xbar{N}} \xbar{G}_{\alpha\beta}\dot{q}^\beta.
\eal
Inverting \eqref{pa} - $\dot{q}^\alpha=\xbar{N}\,\xbar{G}^{\alpha\beta}\pi_\beta$ - and using the Legendre transformation we arrive to (from this point on we omit the bar symbolism, as it must be understood that we are going to work only in the constant potential parametrization)
\bal\label{phHam}
H=\dot{q}^\alpha\pi_\alpha-L\Rightarrow H=N\mathcal{H}, \quad \mathcal{H}=\frac{1}{2}\,G^{\alpha\beta}\pi_\alpha\pi_\beta+1
\eal
where $\mathcal{H}\equiv 0$ is the quadratic constraint. The equations of motion resulting from \eqref{phHam} are
\bal\label{eom}
\dot{q}^\alpha=\{q^\alpha,H\}\Rightarrow \dot{q}^\alpha=N\, G^{\alpha\kappa}\pi_\kappa, \quad
\dot{\pi}_\alpha=\{\pi_\alpha,H\}\Rightarrow \dot{\pi}_\alpha=-\frac{N}{2}\,\ts{G}{^{\kappa\lambda}_{,\alpha}}\pi_\kappa\pi_\lambda
\eal

Let $\xi^\alpha$ be a conformal Killing vector field (CKV) of the supermetric $G_{\alpha\beta}$, i.e.
\bal\label{ckv}
\pound_\xi G_{\alpha\beta}=\omega(q)\,G_{\alpha\beta}\Leftrightarrow \pound_\xi G^{\alpha\beta}=-\omega(q)\,G^{\alpha\beta}
\eal
where the conformal factor $\omega(q)$ is either $\omega(q)\neq \text{constant}$, a \emph{proper} CKV, or $\omega(q)=\text{constant}\neq 0$, a \emph{homothetic} Killing vector field, or $\omega(q)=0$, a \emph{Killing} field.

With the aid of $\xi^\alpha$ we construct the scalar $Q=\xi^\alpha\pi_\alpha$, the evolution of which is
\bal\label{Qdot}
\frac{dQ}{dt}=&\left\{Q,H\right\}&&\non\\
=&\left\{\xi^\alpha\pi_\alpha,N\mathcal{H}\right\}&&\non\\
=&-\frac{1}{2}\,N\left(\pound_\xi G^{\alpha\beta}\right)\pi_\alpha\pi_\beta && \text{where}\,\pound_\xi G^{\alpha\beta}=\xi^\kappa \ts{G}{^{\alpha\beta}_{,\kappa}}- \ts{\xi}{^{\alpha}_{,\kappa}}G^{\kappa\beta}- \ts{\xi}{^{\beta}_{,\kappa}}G^{\alpha\kappa}\non\\
=&+\frac{1}{2}\,N\omega(q)\,G^{\alpha\beta}\pi_\alpha\pi_\beta && \text{by}\, \eqref{ckv}\non\\
=&-N\omega(q) && \text{by}\, \eqref{phHam}\, \text{and}\, \mathcal{H}=0
\eal
For convenience we can choose the time gauge $N\,dt=d\tau$, thus transforming \eqref{Qdot} into
\bal\label{fQdot}
\frac{dQ}{d\tau}=-\omega(q).
\eal

The following cases can be deduced from \eqref{fQdot} depending on the value of $\omega(q)$:
\begin{itemize}
\item $\omega(q)=0$ i.e. the field $\xi^\alpha$ corresponds to a Killing field of $G_{\alpha\beta}$, then \eqref{fQdot} defines an \emph{integral of motion} $Q=c\Rightarrow \xi^\alpha\pi_\alpha=c.$
\item $\omega(q)=-c_\omega$ i.e. the field $\xi^\alpha$ corresponds to a homothetic field of $G_{\alpha\beta}$, then \eqref{fQdot} defines a \emph{semi-integral of motion} $\disfrac{dQ}{d\tau}=c_\omega\Rightarrow \xi^\alpha\pi_\alpha=c_\omega\tau+c.$ The constant $c_\omega$ can always chosen equal to one, by an appropriate rescaling of the field $\xi^\alpha$.
\item $\omega(q)\neq \text{constant}$ i.e. the field $\xi^\alpha$ corresponds to a proper CKV field, then \eqref{fQdot} defines a \emph{relation} between the phase space variables $(q^\alpha,\pi_\alpha)$, which however is nothing but a multiple of the constraint.
\end{itemize}

Indeed in all three cases the relation induced by \eqref{Qdot} is compatible with the equations of motion \eqref{eom}
\bal\label{w relation}
-\omega(q)&=\frac{1}{N}\frac{dQ}{dt}&&\non\\
&=\frac{1}{N}\left(\dot{\xi}^\alpha\pi_\alpha+\xi^\alpha\dot{\pi}_\alpha\right)&&\non\\
&=\frac{1}{N}\left(\ts{\xi}{^\alpha_{,\sigma}}\dot{q}^\sigma\pi_\alpha+\xi^\alpha\dot{\pi}_\alpha\right)\non&&\\
&=\ts{\xi}{^\alpha_{,\sigma}}G^{\sigma\beta}\pi_\beta\pi_\alpha-\frac{1}{2}\,\xi^\alpha\ts{G}{^{\kappa\lambda}_{,\alpha}}\pi_\kappa\pi_\lambda && \text{from}\, \eqref{eom}\non\\
&= \ts{\xi}{^\kappa_{,\sigma}}G^{\sigma\beta}\pi_\kappa\pi_\lambda- \frac{1}{2}\,\left(\pound_\xi G^{\kappa\lambda}+\ts{\xi}{^\kappa_{,\sigma}}G^{\sigma\lambda}+\ts{\xi}{^\lambda_{,\sigma}}G^{\kappa\sigma}\right)\pi_\kappa\pi_\lambda\non\\
&=-\frac{1}{2}\,\left(\pound_\xi G^{\kappa\lambda}\right)\pi_\kappa\pi_\lambda\non\\
&=+\frac{1}{2}\left(\omega(q) G^{\kappa\lambda}\right)\pi_\kappa\pi_\lambda\Rightarrow && \text{by}\, \eqref{ckv}\non\\
0&=\omega(q)\left(\frac{1}{2}\,G^{\kappa\lambda}\pi_\kappa\pi_\lambda+1\right)\Rightarrow\non\\
0&=\omega(q)\mathcal{H}
\eal
which is an identity due to the nihilism of the constraint $\mathcal{H}$. Thus indicating that there are no extra relations among the velocities except the constraint itself, a fact that is welcomed since otherwise the geometry of the configuration space would not be compatible with the dynamics of the system.

The usage of the three above possibilities for the CKV fields can be summarized as follows: The first one, the case of a Killing field, is a well known theorem used in the study of the geodesics of a Riemannian geometry. The second case concerning the homothetic field has been recently discussed in \cite{Tsamp2}, \cite{Tsamp4}. The third case concerning the proper CKV field has not, to the best of our knowledge, been discussed in the context of minisuperspace, although in the case of null geodesics on pseudo-Riemannian manifolds both proper conformal Killing fields and conformal Killing tensors have been considered since they lead to integrals of motions, see \cite{Edgar} and references therein.

The use of the homothetic field is straightforward since it reduces the order of the second set of the equations of motion \eqref{eom} by one.

The usefulness of all the proper CKV's is a kind of trivial one, since \eqref{w relation} can be rewritten as $\frac{d}{dt}\left(Q+\int\!\! N\omega dt\right)=\omega \mathcal{H}$ and with $\omega(q)$ in principle unknown as functions of $t$ one can not perform the integral in order to arrive at a relation between $\dot{q}^\kappa$ and $q^\kappa$. In summary, we conclude that all the conformal Killing fields (proper or not) lead to integrals of motion; the Killing fields produce integrals of motion, the homothetic field a semi-integral of motion while the proper conformal Killing fields describe multiples of the constraint.

\section{Kantowski - Sachs model in vacuum}

Kantowski - Sachs (KS) spacetimes can be defined locally as those
admitting a $G_3$ isometry group acting on two-dimensional space-like orbits
of positive curvature, thus as spherically symmetric. Under the above definition the line element describing KS spacetimes can be written
\bal\label{ds KS}
ds^2=-N(t)^2\,dt^2+a(t)^2\,dr^2+b(t)^2\left(d\theta^2+\sin^2\theta\,d\phi^2\right)
\eal

The Hilbert - Einstein Lagrangian in the case of the above metric, reads
\bal\label{Lgr1}
L_g=-\frac{2}{N(t)}\,a(t)\,b'(t)^2-\frac{4\,b(t)}{N(t)}\,a'(t)\,b'(t)+2\,a(t)\,N(t)
\eal
which can be cast into the form
\bal\label{Lgr2}
L_g=\frac{1}{2\,N}\,G_{\alpha\beta}q'^\alpha q'^\beta-N\,V,\quad G_{\alpha\beta}=\begin{pmatrix}0&-4b\\-4b&-4a\end{pmatrix}\, \text{and} \, V=-2a
\eal
The \emph{physical gauge} is accomplished with the redefinition of the lapse function $N\to \xbar{N}:=N/V$ resulting to the Lagrangian
\bal\label{phLgr}
L_g=\frac{1}{2\,\xbar{N}}\, \xbar{G}_{\alpha\beta}q'^\alpha q'^\beta-\xbar{N}, \quad \xbar{G}_{\alpha\beta}=\begin{pmatrix}0&8\,a\,b\\8\,a\,b&8\,a^2\end{pmatrix}
\eal

It is easy to see that there are infinite CKV fields of the supermetric $\xbar{G}_{\alpha\beta}$, i.e.
\bal\label{conξ1}
\xi=\left(a\,f_1(a^2\,b)-\frac{a}{2b}\,f_2(b)\right)\partial_a+f_2(b)\partial_b
\eal
along with the corresponding conformal factors
\bal\label{ωf1f2}
\omega=2f_1(a^2\,b)+2a^2\,b\,f'_1(a^2\,b)+f'_2(b)
\eal
where $f_1,\,f_2$ are arbitrary functions of their arguments. This arbitrariness is expected due to the configuration space being two dimensional. Since the conformal factor $\omega$ does not have a fixed value, we can search for those $f_1,\, f_2$ which reduce it to a constant, in other words we are searching for the homothetic/Killing vector fields of the supermetric.

To this end, let us set $a^2\,b=u$ and $\omega=c$, in equation \eqref{ωf1f2}
\bal\label{ωc}
2f_1(u)+2u\,f'_1(u)+f'_2(b)=c
\eal
If we differentiate equation \eqref{ωc} with respect to $b$ we have
$f''_2(b)=0\Rightarrow f_2(b)=k_1\,b+k_2$. Insertion of this result into \eqref{ωc} results in the ode $2u\,f'_1(u)+2f_1(u)+k_1-c=0$, whose solution is $f_1(u)=2k_3/u+(c-k_1)/2$. If we collect all the pieces, we have three Killing fields $\xi_{(I)}$ and one homothetic field $h$
\bal\label{homKS}
\xi_{(1)}=-a\,\partial_a+b\,\partial_b,\quad
\xi_{(2)}=-\frac{a}{2b}\,\partial_a+\partial_b,\quad
\xi_{(3)}=\frac{1}{a\,b}\,\partial_a \quad
\text{and}\quad
h=\frac{a}{2}\,\partial_a
\eal
From the phase space point of view where
\be
\pi_a:= \frac{\partial L_g}{\partial a'} = \frac{8\, a\, b\, b'}{\xbar{N}} ,\quad\quad \pi_b:= \frac{\partial L_g}{\partial b'} = \frac{8\,a\,\left(a\, b\right)'}{\xbar{N}} ,
\ee
the corresponding integrals of motion in the gauge $\xbar{N}(t)\,dt=d\tau$ are
\bsub\label{int KS}
\bal
Q_1:=&-a\,\pi_\alpha+b\,\pi_b=c_1\Rightarrow 8\,a(\tau)\,b(\tau)^2\,a'(\tau)=c_1\label{int1 KS}\\
Q_2:=&-\frac{a}{2\,b}\,\pi_\alpha+\pi_b=c_2\Rightarrow 4\,a(\tau)\,\left(2\,b(\tau)\,a'(\tau)+a(\tau)\,b'(\tau)\right)=c_2\label{int2 KS}\\
Q_3:=&\frac{1}{a\,b}\,\pi_\alpha=c_3\Rightarrow 8\,b'(\tau)=c_3\label{int3 KS}\\
Q_4:=&\frac{a}{2}\,\pi_\alpha =-\tau+c_4 \Rightarrow 4\,a(\tau)^2\,b(\tau)\,b'(\tau)= -\tau+c_4\label{int4 KS},
\eal
\esub
where the last integral, \eqref{int4 KS}, was constructed using \eqref{fQdot} with $\omega=1$ being the homothetic factor of the field $h$.
By solving \eqref{int KS} algebraically for $a(\tau)$, $a'(\tau)$, $b(\tau)$, $b'(\tau)$ and applying the consistency conditions $a'(\tau)=\disfrac{da(\tau)}{d\tau},\, b'(\tau)=\disfrac{db(\tau)}{d\tau}$ we finally have
\bal\label{sol KS}
a(\tau)=\pm\frac{c_2}{2}\sqrt{\frac{c_4-\tau}{\tau-c_1-c_4}},\,b(\tau)=\frac{c_1+c_4-\tau}{c_2}\quad \text{and} \quad c_2\,c_3=-8.
\eal
In this gauge, the lapse is $\xbar{N}=1\Rightarrow N = -\frac{1}{2\, a}$, thus the line element \eqref{ds KS} can be written as
\begin{align} \nonumber
ds^2 = &-\frac{\tau-c_1-c_4}{c_2^2 \,(c_4-\tau)} d\tau^2 + \frac{c_2^2\,(c_4-\tau)}{4\,(\tau-c_1-c_4)}\, dr^2 + \left(\frac{\tau-c_1-c_4}{c_2}\right)^2 d\theta^2 + \\ \label{lineel1}
&+\left(\frac{\tau-c_1-c_4}{c_2}\right)^2\sin^2\theta\, d\phi^2,
\end{align}
which, of course, is the Lorentzian solution to Einstein's vacuum equations $R_{\mu\nu}=0$, first reported in \cite{Kant}. Line element \eqref{lineel1} can be further simplified by the following consecutive transformations (that result in discarding the non essential constants appearing in it): (a) a time translation $\tau\rightarrow \tau+ c_4$, (b) a scaling $(\tau,\,r,\,\phi)\rightarrow (c_2 \, \tau,\frac{2\,r}{c_2},\,c_2\,\phi)$, (c) a redefinition $c_1=c_2\, c$ and (d) a final translation $\tau\rightarrow c-\tau$. Under these changes metric \eqref{lineel1} becomes
\begin{equation}\label{linefin1}
ds^2 = -\frac{\tau}{c-\tau} d\tau^2 + \frac{c-\tau}{\tau}\, dr^2 +\tau^2 d\theta^2 + \tau^2\sin^2\theta\, d\phi^2,
\end{equation}
which verifies that the Kantowski - Sachs geometry has one essential constant, as it is expected.

It is interesting that solutions of Euclidean and/or neutral signature can also be acquired. To this end we adopt the gauge $\xbar{N}(t)\,dt=\ima\, d\tau$, in which case the symmetries \eqref{homKS} lead to the integrals of motion
\bsub\label{int KS2}
\bal
Q_1:=&-a\,\pi_\alpha+b\,\pi_b=\tilde{c}_1\Rightarrow -8\,\ima\,a(\tau)\,b(\tau)^2\,a'(\tau)=\tilde{c}_1\label{int1 KS2}\\
Q_2:=&-\frac{a}{2\,b}\,\pi_\alpha+\pi_b=\tilde{c}_2\Rightarrow \nonumber \\ & -4\,\ima\,a(\tau)\,\left(2\,b(\tau)\,a'(\tau)+a(\tau)\,b'(\tau)\right)=c_2\label{int2 KS2}\\
Q_3:=&\frac{1}{a\,b}\,\pi_\alpha=\tilde{c}_3\Rightarrow -8\,\ima\,b'(\tau)=\tilde{c}_3\label{int3 KS2}\\
Q_4:=&\frac{a}{2}\,\pi_\alpha=-\ima\, \tau+\tilde{c}_4\Rightarrow -4\,\ima\,a(\tau)^2\,b(\tau)\,b'(\tau)= -\ima\, \tau+\tilde{c}_4\label{int4 KS2}.
\eal
\esub
In order to simplify the resulting solution, we choose the constants to be $\tilde{c}_i = \ima\, c_i$, for $i=1,\ldots,4$. Once more the system \eqref{int KS2} can be solved algebraically for $a(\tau)$, $a'(\tau)$, $b(\tau)$ and $b'(\tau)$; this solution, together with the consistency conditions $a'(\tau)=\disfrac{da(\tau)}{d\tau}$ and $b'(\tau)=\disfrac{db(\tau)}{d\tau}$ leads to
\bal\label{sol KS2}
a(\tau)=\pm\frac{c_2}{2}\sqrt{\frac{c_4-\tau}{\tau-c_1-c_4}},\,b(\tau)=\frac{c_1+c_4-\tau}{c_2}\quad \text{and} \quad c_2\,c_3=8.
\eal
The lapse function becomes $N=- \frac{\ima}{2\,a}$ and the corresponding line element reads
\begin{align} \nonumber
ds^2 = &\frac{c_1+c_4-\tau}{c_2^2 \,(c_4-\tau)} d\tau^2 + \frac{c_2^2\,(c_4-\tau)}{4\,(c_1+c_4-\tau)}\, dr^2 + \left(\frac{\tau-c_1-c_4}{c_2}\right)^2 d\theta^2 + \\ \label{lineel2}
&+\left(\frac{\tau-c_1-c_4}{c_2}\right)^2\sin^2\theta\, d\phi^2.
\end{align}
We can use the same transformations as before to clear \eqref{lineel2} from the non essential constants obtaining the final line element
\begin{equation} \label{linefin2}
ds^2 = \frac{\tau}{\tau-c} d\tau^2 + \frac{\tau-c}{\tau}\, dr^2 + \tau^2 d\theta^2 + \tau^2\sin^2\theta\, d\phi^2.
\end{equation}
This line element, depending on the range of $\tau$ with respect to the essential constant $c$, describes a solution of either Euclidean (first obtained in a different way in \cite{Lorentz}) or neutral signature. Note that \eqref{linefin1} and \eqref{linefin2} develop a curvature singularity at $t=0$, since the Kretschmann scalar is $\frac{12\, c^2}{\tau^6}$.

\section{Discussion}

In this paper we consider the theory of symmetries of coupled differential equations concerning singular systems in the case of minisuperspace framework. The new ingredient in our analysis is that we let the action of the symmetry generators on the Lagrangian and/or to the equations of motions to equal a multiply of the constraint. Moreover we \emph{do not} fix the gauge, thus treating the lapse function $N(t)$ as a dynamical degree of freedom.

The results of the above analysis are:
\begin{itemize}
\item The variational symmetries of the action \eqref{act} are described by the simultaneous conformal Killing fields of the metric $G_{\mu\nu}$ and of the potential $V$, with opposite conformal Killing factors, along with the time reparametrisation symmetry. The former ones, are exactly the conditional symmetries found in our earlier work \cite{tchris1} in the context of phase space.
\item The Lie - point symmetries of the Euler - Lagrange equations emanating from \eqref{act} are the variational symmetries plus the scaling symmetry. In detail the  resulting symmetries are: (a) the simultaneous conformal Killing fields $\eqref{xiunder}$ entering $X_1$ \eqref{gen1}, (b) the reparametrisation generator $X_2$ \eqref{gen2} and (c) the well known scaling symmetry generator $Y$ \eqref{genY}. The latter two are the specialization of the corresponding already known generators from the full Einstein gravity theory (see e.g. \cite{Stephani} pp 158-159). The generator $X_1$ encompasses the information regarding the combined symmetries of the minisuperspace metric and the potential. The case of the nonconstant conformal factor $\underline{\tau}(q)$ is not previously presented in the literature. In the particular parametrisation of the lapse in which the potential becomes constant, the symmetries $X_1$, $Y$ are transformed into the Killing and the homothetic symmetries of the scaled minisuperspace metric respectively; a fact that establishes the connection to the known symmetries.

\end{itemize}

The benefit of this perspective is that one can make contact between the variational and the Lie - point symmetries. If one chooses to apply the standard procedure for finding the Lie - point symmetries (i.e. the one for regular systems) then he is forced to demand $pr^{(2)}X(E^\kappa)=0$ instead of \eqref{newcrit2}. In this case the resulting symmetries are
\bsub\label{regsym}
\bal
\pound_\xi \Gamma^\kappa_{\mu\nu}&=\frac{1}{2}\,\left(h_{,\mu}\delta^\kappa_\nu+h_{,\nu}\delta^\kappa_\mu\right)\label{pc}\\
\pound_\xi V^{,\kappa}&=-2h\,V^{,\kappa}\label{resV}
\eal
\esub
i.e. the projective collineations \eqref{pc} of the connection $\Gamma^\kappa_{\mu\nu}$, along with a restriction \eqref{resV} on the components of the $\xi^\alpha$. It is interesting to notice that the Lagrangian \eqref{Lag} can be used to describe the geodesics problem in a Riemmanian space when we set $N=1$ and  $V=0$. This special case was studied in \cite{Tsamp1} and the result was the first of \eqref{regsym}, with complete agreement with our general results.

If on the other hand, one chooses to fix the gauge, then the constraint is lost since there is no variation of the lapse function $N(t)$. Furthermore the gauge fixing of the lapse may lead to a loss of symmetries (see appendix of \cite{tchris1} for an example). It is noteworthy that, even if one has gauge fixed the lapse, the complete set of symmetries can be acquired by allowing the action of the generator to produce a multiple of the constraint.

A natural question about the modification of infinitesimal criterion for Lie symmetries, i.e. equations \eqref{newcrit1}, \eqref{newcrit2} is why we do not modify the corresponding criterion \eqref{cr1}. The answer lies to the fact that the action of the part $\omega\partial_N$ of the generator \eqref{generator} on \eqref{act} reproduces the constraint \eqref{con} multiplied by $\omega(t,q,N)$, thus the constraint is already embedded in that (standard) criterion.

In order to make the whole discussion work into practice we employ our results in the case of the vacuum Kantowski - Sachs spacetime, obtaining the known classical solution of Kantowski - Sachs \cite{Kant}, along with the Euclidean solution obtained by D. Lorentz in \cite{Lorentz}. It is also noteworthy that in our approach it was not necessary to solve the corresponding Einstein's equations which are of second order; we only needed to solve the integrals of motion and the symmetry equations on the configuration space which all are of first order.

Finally, as a by product of our analysis, we have found that the maximum number of Lie - point symmetries for the minisuperspace models is $\disfrac{n(n+1)}{2}+2$.

\pagebreak
\appendix
\section{Appendix A: Terms of equation \eqref{cr1}} \label{app1}
\textbf{Terms from the action of $pr^{(1)}X$ on $L$}
\begin{align}
\chi \frac{\partial L}{\partial t} & =0\\
\xi^\alpha \frac{\partial L}{\partial q^\alpha}& = \frac{\xi^\alpha}{2N}G_{\kappa\lambda,\alpha}\dot{q}^\kappa\dot{q}^\lambda- N\xi^\alpha V_{,\alpha} \\
\omega \frac{\partial L}{\partial N} &= -\frac{\omega}{2N^2}G_{\kappa\lambda}\dot{q}^\kappa\dot{q}^\lambda - \omega V \\ \nonumber
\phi^\alpha \frac{\partial L}{\partial \dot{q}^\alpha}& = \frac{\xi^\alpha_{,t}}{N}G_{\kappa\alpha}\dot{q}^\kappa - \frac{\chi_{,t}}{N}G_{\kappa\alpha}\dot{q}^\alpha\dot{q}^\kappa+ \left(\frac{\xi^\alpha_{,\lambda}}{2 N}G_{\kappa\alpha}\dot{q}^\kappa\dot{q}^\lambda+ \frac{\xi^\alpha_{,\kappa}}{2 N} G_{\alpha\lambda}\dot{q}^\kappa\dot{q}^\lambda\right)- \\
& - \frac{\chi_{,\beta}}{N} G_{\kappa\alpha}\dot{q}^\kappa\dot{q}^\alpha\dot{q}^\beta+ \frac{\dot{N}}{N}\xi^\alpha_{,0}G_{\kappa\alpha}\dot{q}^\kappa- \frac{\dot{N}}{N}\chi_{,0}G_{\kappa\alpha}\dot{q}^\kappa\dot{q}^\alpha
\end{align}
\textbf{Terms from $\frac{d\chi}{dt}\, L$ and $\frac{df}{dt}$.}
\begin{align} \nonumber
\frac{d\chi}{dt}\, L & =\left(\chi_{,t}+\dot{q}^\alpha \chi_{,\alpha}+ \dot{N}\chi_{,0}\right) \left(\frac{1}{2N}G_{\kappa\lambda}\dot{q}^\kappa\dot{q}^\lambda-N V\right)  \\  \nonumber
& =\frac{\chi_{,t}}{2N}G_{\kappa\lambda}\dot{q}^\kappa\dot{q}^\lambda - \chi_{,t}NV+ \frac{\chi_{,\alpha}}{2N}G_{\kappa\lambda}\dot{q}^\kappa\dot{q}^\lambda\dot{q}^\alpha - \chi_{,\alpha} N V\dot{q}^\alpha \\
&\ph{=}+ \frac{\dot{N}}{2N}\chi_{,0}G_{\kappa\lambda}\dot{q}^\kappa\dot{q}^\lambda + \dot{N} N \chi_{,0} V\\
\frac{d f}{dt}&= f_{,t}+\dot{q}^\alpha f_{,\alpha}+\dot{N}f_{,0}
\end{align}

\section{Appendix B: The action of $pr^{(2)}X$ on $E^\kappa$} \label{app2}
\begin{align}
\chi \frac{\partial E^\kappa}{\partial t} &=0 \\
\xi^\alpha \frac{\partial E^\kappa}{\partial q^\alpha} &= \xi^\alpha \Gamma^\kappa_{\mu\nu,\alpha}\dot{q}^\mu\dot{q}^\nu+ N^2 \xi^\alpha V^\kappa_{,\alpha} \\
\omega \frac{\partial E^\kappa}{\partial N} &=  \frac{\omega}{N^2} \dot{N}\dot{q}^\kappa+2\omega N V^{,\kappa}\\
\phi^\alpha\frac{\partial E^\kappa}{\partial \dot{q}^\alpha} &= 2\, \xi^\alpha_{,t}\, \Gamma^\kappa_{\alpha\mu}\,  \dot{q}^\mu - \frac{\xi^\kappa_{,t}}{N}\, \dot{N} \\ \nonumber
& +\left(\left(\xi^\alpha_{,\mu}-\chi_{,t}\delta^\alpha_\mu\right)\Gamma^\kappa_{\alpha\nu} + \left(\xi^\alpha_{,\nu}-\chi_{,t}\delta^\alpha_\nu\right)\Gamma^\kappa_{\alpha\mu} + \frac{1}{2N}\left(\chi_{,\mu}\delta^\kappa_\nu+ \chi_{,\nu}\delta^\kappa_{\mu}\right) \right) \dot{q}^\mu\dot{q}^\nu  \\ \nonumber
&+ \left(2\xi^\alpha_{,0}\, \Gamma^\kappa_{\alpha\beta}-\frac{1}{N}\left(\xi^\kappa_{,\beta}- \chi_{,t}\delta^\kappa_{\beta}\right)\right)\dot{N}\dot{q}^\beta- 2 \, \chi_{,(\beta}\, \Gamma^\kappa_{\alpha\nu)} \, \dot{q}^\alpha\dot{q}^\beta\dot{q}^\nu \\
& - \frac{\xi^\kappa_{,0}}{N} \dot{N}^2- 2\chi_{,0}\, \Gamma^\kappa_{\mu\nu} \dot{N} \dot{q}^\mu \dot{q}^\nu + \frac{\chi_{,0}}{N} \dot{N}^2 \dot{q}^\kappa \\
\Omega \frac{\partial E^\kappa}{\partial \dot{N}} &= -\frac{\omega_{,t}}{N}\dot{q}^\kappa- \frac{(\omega_{,0}-\chi_{,t})}{N}\dot{N}\dot{q}^\kappa- \frac{\omega_{,\beta}}{N}\dot{q}^\beta\dot{q}^\kappa+ \frac{\chi_{,\beta}}{N} \dot{N}\dot{q}^\beta\dot{q}^\kappa+ \frac{\chi_{,0}}{N}\dot{N}^2\dot{q}^\kappa \\
\Phi^\alpha \frac{\partial E^\kappa}{\partial \ddot{q}^\alpha}& = \Phi^\kappa
\end{align}

\section{Appendix C: Proof of $h(q)=\tau(q)+c$} \label{proof}
We begin from the relation
\begin{align} \nonumber
V^{,\alpha}=G^{\alpha\mu}V_{,\mu} \Rightarrow \pound_\xi V^{,\alpha}& = (\pound_\xi G^{\alpha\mu})V_{,\mu} + G^{\alpha\mu} \pound_\xi V_{,\mu} \\ \nonumber
\Rightarrow \pound_\xi V^{,\alpha} &= -\tau G^{\mu\alpha} V_{,\mu}+G^{\alpha\mu} \pound_\xi V_{,\mu} \Rightarrow \\ \label{C1}
\Rightarrow \pound_\xi V^{,\alpha} &=-\tau G^{\mu\alpha} V_{,\mu} + G^{\alpha\mu}\left(\xi^\rho V_{,\mu,\rho}+\xi^\rho_{,\mu} V_{,\rho}\right)
\end{align}
where \eqref{confg} has been used. If we write
\be
\xi^\rho V_{,\rho,\mu} = (\xi^\rho V_{,\rho})_{,\mu} - \xi^\rho_{,\mu} V_{,\rho}=(\pound_\xi V)_{,\mu} - \xi^\rho_{,\mu} V_{,\rho} \, ,
\ee
insert this result into \eqref{C1} and use \eqref{confv} we get
\begin{align} \nonumber
\pound_\xi V^{,\alpha} &=-\tau G^{\mu\alpha} V_{,\mu} + G^{\alpha\mu} (\pound_\xi V)_{,\mu}= \\ \nonumber
 &=-\tau G^{\mu\alpha} V_{,\mu}+ G^{\alpha\mu}\left((\tau - 2 h)V\right)_{,\mu} \Rightarrow \\ \label{C2}
 \pound_\xi V^{,\alpha} &=G^{\alpha\mu}\tau_{,\mu} V - 2G^{\alpha\mu} h_{,\mu} V - 2 G^{\alpha\mu} h V_{,\mu}.
\end{align}
If we insert \eqref{C2} into \eqref{pkap}, $P^\kappa$ becomes
\be \label{C4}
P^\kappa = \frac{1}{2}\tau^{,\kappa}-h^{,\kappa}
\ee
Substitution of \eqref{C4} and \eqref{confgamma} into \eqref{liegam} results in
\be \label{C3}
(\tau-h)_{,\mu} \delta^\kappa_\nu + (\tau-h)_{,\nu}\delta^\kappa_\mu - 2 G_{\mu\nu}G^{\kappa\rho}(\tau-h)_{,\rho} =0.
\ee
If we set $A(q)=\tau(q)-h(q)$ and contract $\kappa$ with $\nu$ we are led to
\be
(n-1) A_{,\mu}=0
\ee
where $n$ is the dimension of the supermetric $G_{\alpha\beta}$. Thus, for $n> 1$
\be
A_{,\mu}=0 \Rightarrow h(q) = \tau(q) +c
\ee


\begin{thebibliography}{99}

\bibitem{SLie}  Sophus Lie, Mathematische Annalen \textbf{16}, Issue 4, (1880) 441-528

\bibitem{Ibra} Nail H. Ibragimov, \textit{Selected works, vol I, II, III, and IV}, ALGA Publications Blekinge Institute of Technology Karlskrona, Sweden (2006)

\bibitem{Stephani} Stephani, H. \textit{Differential Equations: Their solution using symmetries}, Cambridge University Press, Cambridge (1989)

\bibitem{Olver1} Olver, P.J., \textit{Equivalence, Invariants and Symmetry}, Cambridge University Press, Cambridge (1995)

\bibitem{Olver2} Olver, P.J., \textit{Applications of Lie Groups to Differential Equations}, 2nd ed., Springer - Verlag, Berlin, Heidelberg, New York (2000)

\bibitem{Marmo} S. Capozziello, G. Marmo, C. Rubano, P. Scudellaro, Int. J. Mod. Phys. D \textbf{6}, 4 (1997) 491-503.

\bibitem{Cotsakis1} S. Cotsakis, P.G.L. Leach, H. Pantazi Grav. Cosmol. \textbf{4} (1998) 314-325

\bibitem{Lamb1} S. Capozziello, G. Lambiase, Gen. Rel. Grav. \textbf{32} (2000) 295-311

\bibitem{Lamb2} S. Capozziello, G. Lambiase, Gen. Rel. Grav. \textbf{32} (2000) 673-696

\bibitem{Cotsakis2} P.G.L. Leach, S. Cotsakis, J. Miritzis, Grav. Cosmol. \textbf{7} (2001) 311-320 and references therein

\bibitem{Vakili} B. Vakili and F. Khazaie, Class. Quantum Grav. \textbf{29} (2012) 035015

\bibitem{Falciano} F. T. Falciano, Roberto Pereira, N. Pinto-Neto and E. Sergio Santini, Phys.Rev. \textbf{D86} (2012) 063504

\bibitem{Capoz} S. Capozziello, M. De Laurentis and S.D. Odintsov, Eur.Phys.J. \textbf{C72} (2012) 2068

\bibitem{Sarkar} K. Sarkar, Nayem Sk., S. Debnath, and A. K. Sanyal, arXiv: 1207.3219v1 [astro-ph.CO] 13 July 2012

\bibitem{Tsamp1} M. Tsamparlis and A. Paliathanasis, Gen. Rel. Grav. \textbf{42} (2010) 2957-2980

\bibitem{Tsamp2} M. Tsamparlis and A. Paliathanasis, J. Phys. A \textbf{A44} (2011) 175202

\bibitem{Tsamp3} M. Tsamparlis and A. Paliathanasis, Gen. Rel. Grav. \textbf{43} (2011) 1861-1881

\bibitem{tchris2} T. Christodoulakis and Petros A. Terzis, J. Math. Phys. \textbf{47} (2006) 102502

\bibitem{tchris3} T. Christodoulakis and Petros A. Terzis, Class. Quantum Grav. \textbf{24} (2007) 875-887

\bibitem{tchris4} Petros A. Terzis and T. Christodoulakis, Gen. Rel. Grav. \textbf{41} (2009) 469-495

\bibitem{tchris5} Petros A. Terzis and T. Christodoulakis, Class. Quantum Grav. \textbf{29} (2012) 235007

\bibitem{Dewitt} B. S. DeWitt, in \emph{Relativity}, Carmeli et al., eds. Plenum, New York (1970) pp. 359 - 374

\bibitem{Misner} C. W. Misner, in \emph{Magic Without Magic: John Archibald Wheeler}, W. H. Freeman and company, San Francisco (1972) pp. 441 - 473

\bibitem{tchris1} T. Christodoulakis, N. Dimakis, Petros A. Terzis, G. Doulis, Th. Grammenos, E. Melas and A. Spanou, \emph{Preprint} gr-qc/1208.0462 (2012)

\bibitem{Kuchar} K.V. K\v uchar, J. Math. Phys. \textbf{23} (1982) 1647 - 1661

\bibitem{Tsamp4} A. Paliathanasis, M. Tsamparlis, S. Basilakos, Phys. Rev. D \textbf{84} (2011) 123514 

\bibitem{Edgar}  R. Rani, S. Brian Edgar and A. Barnes,  Class. Quant. Grav. \textbf{20} (2003) 1929-1942

\bibitem{Kant} R. Kantowski, R. K. Sachs. J.Math.Phys. \textbf{7} (1966) 443

\bibitem{Lorentz} D. Lorentz, Acta Physica Polonica, \textbf{B14}, no 11, (1983) 787-789

\end{thebibliography}
\end{document}